\documentclass{article}
\usepackage{waspaa21,amsmath,graphicx,url,times}
\usepackage{caption, subcaption, amsfonts, bm, amssymb, multirow, hyperref}
\usepackage{color}

\input{def.set}

\title{Test-Time Adaptation Toward Personalized Speech Enhancement:\\Zero-Shot Learning with Knowledge Distillation}

\name{Sunwoo Kim, Minje Kim\sthanks{This material is based upon work supported by the National Science Foundation under Grant No. 2046963.}
}
\address{Indiana University, Department of Intelligent Systems Engineering, Bloomington, IN 47408, USA\\ 
\texttt{kimsunw@iu.edu, minje@indiana.edu}\\              
}

\begin{document}

\ninept
\maketitle

\begin{sloppy}

\begin{abstract}
In realistic speech enhancement settings for end-user devices, we often encounter only a few speakers and noise types that tend to reoccur in the specific acoustic environment. We propose a novel personalized speech enhancement method to adapt a compact denoising model to the test-time specificity. Our goal in this test-time adaptation is to utilize no clean speech target of the test speaker, thus fulfilling the requirement for zero-shot learning. To complement the lack of clean utterance, we employ the knowledge distillation framework. Instead of the missing clean utterance target, we distill the more advanced denoising results from an overly large teacher model, and use it as the pseudo target to train the small student model. This zero-shot learning procedure circumvents the process of collecting users' clean speech, a process that users are reluctant to comply due to privacy concerns and technical difficulty of recording clean voice. Experiments  on various test-time conditions show that the proposed personalization method  achieves significant performance gains compared to larger baseline networks trained from a large speaker- and noise-agnostic datasets. In addition, since the compact personalized models can outperform larger general-purpose models, we claim that the proposed method performs model compression with no loss of denoising performance. 
\end{abstract}


\section{Introduction}
\label{sec:intro}

Recent advances in deep learning-based speech enhancement (SE) models have shown superior performance with respect to traditional machine learning and signal processing methods \cite{WangDL2018ieeeacmaslp,XuY2014ieeespl,ChazanSE2017MoE}. These large models are typically trained from a large training set, so they generalize well to various test-time conditions including different speakers, noises, and signal-to-noise ratios (SNR) of the added noise. However, the growing size of neural network architectures and computational complexity renders them difficult to deploy onto resource-constrained devices. Hence, model compression methods have been gaining interest to facilitate the practicality of deep-learning architectures in real-time applications. Some main modes of compression such as quantization, pruning, and knowledge distillation (KD) have shown great promise in dramatically reducing the complexity 
\cite{HanS2016iclr}. 
% \cite{HanS2016iclr,FrankleJ2018iclr}. 
However, these \textit{context-agnostic} compression methods are designed to reduce a general-purpose SE model's complexity without knowing the test-time context that the model will be situated in. Hence, a loss in overall performance and generalization power is inevitable. 

% While the drop in generalization performance is common when there is a mismatch in the training and test data, a large model learned from a large training dataset can minimize the gap. Hence, the tradeoff between performance and model compression is a common challenge in designing deep learning models. 

In some practical use cases though, e.g., a family-owned smart assistant device sitting in the living room, it suffices for the enhancement model to perform well only for the specific test environment. Hence, a \textit{context-aware} fine-tuning method is promising, as it can turn the general-purpose SE model, \textit{the generalist}, into a special-purpose version, \textit{the specialist}. It can be seen as a test-time adaptation to the specific speakers and their acoustic context,  overcoming the generalization losses. We call this kind of fine-tuned specialists \textit{personalized speech enhancement} (PSE) systems.

The topic of domain adaptation has been an active area of research in computer vision, and speech and audio applications as well. One common procedure for domain transfer between different datasets is regularizing the differences between the learned representations of source and target data, and it has been applied for emotion, speech, and speaker recognition \cite{DengJ014ieeespl,SunS2017neucom}. However, these applications were provided ample target data, which cannot be assumed for usual cases. Other methods rely on few-shot adaptation in cases where a small number of ground-truth signals are available \cite{SivaramanA2021selfsupervisedpse}. However, it can be challenging to obtain user information due to recent privacy infringement, data leakage issues and advancement in DeepFake technology rendering customers uneasy towards releasing personal information.
% \cite{rochford2019accessibility}.
With user compliance, the user enrollment phase can obtain trigger phrases from  the users, but these recordings can be contaminated with existing background noise and might not be long enough. 

In contrast to aforementioned approaches, zero-shot learning (ZSL) is a data-free solution suitable for training tasks where no additional labeled data is available  \cite{WangW2019acmtist,XianY2018zeroshotlearning}. In the context of PSE, ZSL is a solution that does not require users' clean speech data or their home acoustic environment, while its goal is still to adapt to the test-time specificity. ZSL is an active research topic for classification tasks, where ZSL frameworks typically infer test-time labels for domain adaptation by extracting and utilizing auxiliary information
\cite{LarochelleH2008aaai,PalatucciM2009nips,RomeraParedesB2015icml,SocherR2013nips}
% \cite{LarochelleH2008aaai,RohrbachM2011cvpr,LiuS2018nips,PalatucciM2009nips,KodirovE2017cvpr,RomeraParedesB2015icml,SocherR2013nips}
. Similarly, ZSL in the speech and audio classification applications extracts semantic properties or articulatory distribution to obtain labels during test-time \cite{DauphinY2013iclr,ChoiJ2019ismir,LiX2020aaai}. However, ZSL for speech enhancement has not been widely studied. In \cite{SivaramanA2020interspeech}, a mixture of local expert model was introduced as a ZSL solution to test-time adaptation of an SE model. It achieves the adaptation goal by employing a classifier to select the most suitable one out of pre-trained specialist models for a given noisy test signal. Although it is a valid adaptation method, it only works on a few pre-defined contexts, i.e., varying signal-to-noise ratio (SNR) and gender of the speaker, rather than adapting to the test-time speaker's personality or the unique context.

In this paper, we present a zero-shot learning approach to PSE, based on the KD framework \cite{HintonG2015arxiv}. As a ZSL method, it does not ask for private signals from the user, while it can still adapt to the user's speech and recording environment, thus qualifying as a PSE method. When ZSL is implemented via KD strategies, it is common to use data synthesis techniques through generative adversarial frameworks where the generator generates fake samples
\cite{ChenH2019iccv,MicaelliP2019nips}
% \cite{ChenH2019iccv,MicaelliP2019nips,YeJ2020cvpr,YooJ2019nips}
or through KD using activation or output statistics from pre-trained teacher models to synthesize pseudo samples \cite{LopesRG2017nips,NayakGK2019icml}. 
% , among many others \cite{nayak2021effectiveness, chawla2021data}. 
% However, training GANs and extracting information from teacher models are both non-trivial tasks. 
Instead of including an intermediate data synthesis step, our proposed model directly uses the teacher model's outputs as ground-truth targets to optimize the student model, where the teacher model is defined by a large generalist model trained from a large training dataset, while the student model is a relatively smaller model, thus fulfilling the efficiency criteria. The basic assumption is that the teacher model's large computational capacity guarantees the generalization goal, i.e., it works well in most test-time environments, whose excellent SE results can be considered as if they were the target clean speech from the student model's perspective.  
% It resembles a KD based domain adaptation framework which adapts acoustic models by minimizing the KL divergence between the output distribution of teacher and student, except the models are fed in different data (clean vs. noisy) and does not concern a zero-shot scenario \cite{li2017large}. 
When deploying our framework, we ultimately only use the student model on the device. The teacher model is envisioned to be placed externally on a cloud server, where the actual fine-tuning operations are conducted. The student models can be frequently updated on the server side and transferred to the user device.  To our best knowledge, this ZSL PSE framework is novel in the field of speech enhancement. 
% , and can greatly free up space and expand the utility of small scale models.

% The contributions of our paper are as follows. First, we propose a zero-shot KD framework for domain adaptation during inference in a privacy-perserving manner. Second, we present a zero-shot framework as a form of model compression by fine-tuning small scale student models to outperform larger generalist models. Third, we provide a systematic evaluation of architecture sizes of the student model in relation to the teacher model's. Finally, we evaluate ConvTasNet \cite{LuoY2019conv-tasnet} architecture for the teacher model to provide a comparison between different variants of teacher architectures (recurrent vs. convolutional) and show our framework is agnostic to the teacher's network architecture.

\begin{figure}
    \centering
    \includegraphics[width=.8\columnwidth]{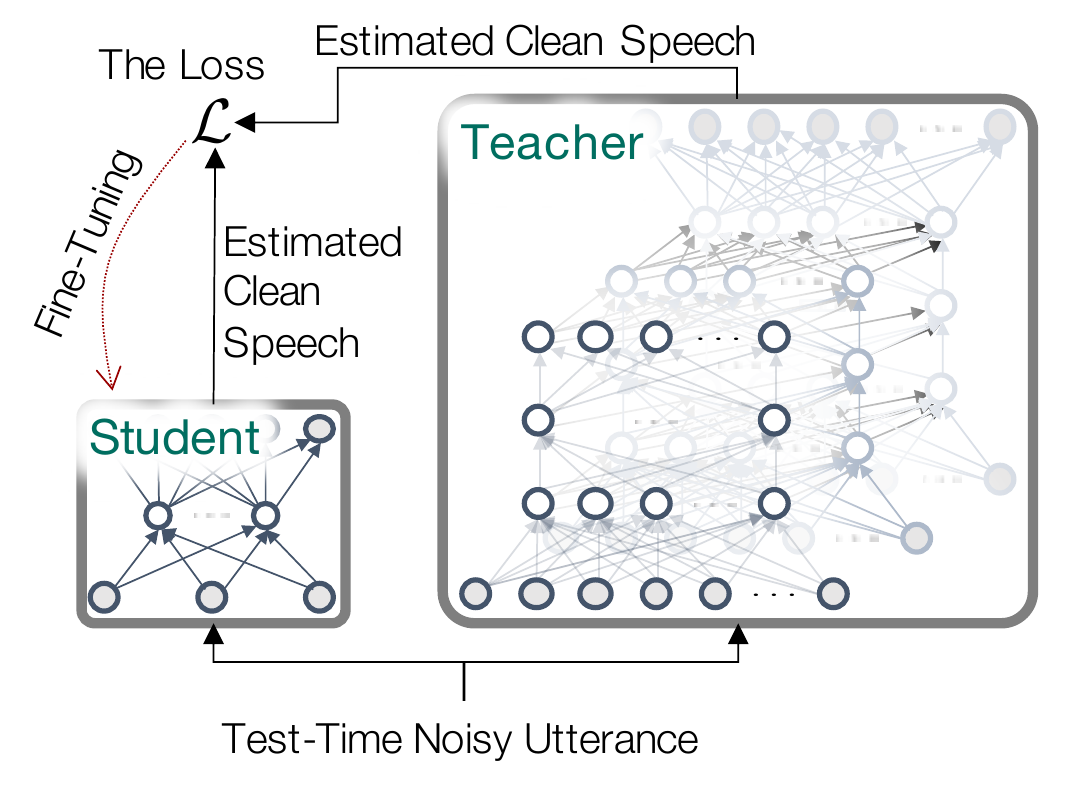}
    \caption{An overview of the proposed KD-based PSE process. The estimated clean speech by the student model is compared against the result from a larger teacher model, whose discrepancy is used to fine-tune the student model. This can be done during the test time as it does not require the clean speech target.}
    \label{fig:overview}
\end{figure}

\section{The Proposed Zero-Shot Learning Method for Personalized Speech Enhancement }
\label{sec:methods}

We implement the proposed ZSL-based PSE via KD. Our goal is to fine-tune a compact student model during the test time, so it adapts to the unseen test speaker and environment. KD plays a key role in our ZSL framework, as its teacher model provides a \textit{pseudo} target for the student model to learn from, while the target clean speech of the test-time noisy utterance is absent. We claim that the proposed PSE method will be helpful when the system needs to deal with the peculiarity of the test-time conditions. This kind of flexibility will be also advantageous for SE models if the system has to be frequently relocated to different test environments. Figure \ref{fig:overview} describes the KD-based PSE process that can fine-tune the student model during the test time. 

\subsection{Pre-training Speech Enhancement Models}
\label{sec:se}

First, we pre-train both the teacher and student models $\calT(\cdot)$ and $\calS(\cdot)$ using a large-scale speech corpus and noise dataset. Here, the teacher model $\calT(\cdot)$ is defined with a large model architecture, so it can properly approximate the complex general-purpose speech denoising function. Meanwhile, even though $\calS(\cdot)$ is trained using the same data, its small capacity hinders it from generalizing well to the unseen test condition. The goal of test-time PSE is to reduce this gap, which will be explained in detail in Sec. \ref{sec:ft}. Note that $\calT(\cdot)$ is not fine-tuned, assuming that its performance as a generalist meets the quality standard in most test cases. Conversely, pre-training $\calT(\cdot)$ can prepare the student model better than a random initialization.

The pre-training phase of the SE models is formulated as follows. We assume an additive signal model where the observed signal $\bm{x}$ is a mixture of a clean speech source $\bm{s}$ and noise source $\bm{n}$ of identical duration: $\bm{x} = \bm{s} + \bm{n}$, which are all monaural.
%for a discrete time index $t$. 
The clean speech utterances are taken from a large dataset containing many speakers, $\bm{s}\in\mathbb{G}$, and the noise recordings are similarly from a large dataset containing various noise types, $\bm{n}\in\mathbb{N}$. The objective is to denoise $\bm{x}$ and estimate waveform $\hat{\bm{s}}$ that closely approximates the target clean speech, e.g., $\bs\approx\hat\bs\leftarrow\calT(x)$. 

% We train SE models under a time-domain loss function based on the scale-invariant signal-to-distortion ratio (SI-SDR) metric expressed in decibels (dB) \cite{LeRouxJL2018sisdr}. We adopt SI-SDR to measure the fidelity of the estimated waveform to the ground-truth target, defined as: 
% \begin{equation}
%     \label{eqn:sisdr}
%     \text{SI-SDR}(\bm{s},\hat{\bm{s}}) = 10\;\text{log}_{10} \frac{||\alpha \bm{s}||^2}{||\alpha \bm{s} - \hat{\bm{s}}||^2}
% \end{equation}
% where both $\bm{s}$ and $\hat{\bm{s}}$ have the same duration and $\alpha=\frac{\hat{\bm{s}}^\top \bm{s}}{||\bm{s}||^2}$ is the optimal scaling factor for the target signal. The generalist SE model is trained to optimize the SI-SDR improvement over the mixtures: 
% \begin{equation}
%     \label{eqn:pre}
%     \argmax_{\bm{\Theta}_G}\; \text{SI-SDR}(\bm{s},\hat{\bm{s}}) - \text{SI-SDR}(\bm{s},\bm{x})
% \end{equation}

The optimization on $\calT(\cdot)$ reduces the loss between the target utterance $\bs$ and reconstruction $\hat\bs$, e.g., $\argmin_{\bTheta_\calT}  \calL(\bs||\calT(\bx; \bTheta_\calT))$,
where $\bTheta_\calT$ denotes the trainable parameters of the teacher model.  Ditto for $\calS(\bx; \bTheta_\calS)$.

% The general-purpose SE models are trained under a vast variety of both speakers and noises to generalize well to test-time environments. 

\subsection{Test-time Personalized Speech Enhancement}
\label{sec:ft}

During the test time, we assume a noisy environment where the SE system is exposed to mixture signals composed of clean speech utterances from the test speaker, $\bm{s}\in\mathbb{S}$, and background noise sources, $\bm{n}\in\mathbb{M}$. For example, in the most extreme case, small generalist models, such as our student models, can fail to generalize well to the test mixtures if those test time sources have not been exposed to the training process, i.e., $\mathbb{G}\cap\mathbb{S}=\varnothing$ and $\mathbb{N}\cap\mathbb{M}=\varnothing$. 

Given these assumptions, we propose a PSE framework that can adapt to a new environment without requiring test user's ground-truth clean speech samples or any other auxiliary information of the speakers and acoustic scene. Since we formulate the proposed PSE method as a fine-tuning process, we begin with a compact student model, $\calS(\cdot)$, pre-trained in a context-agnostic manner as in Sec. \ref{sec:se}. To fine-tune it, its denoising result, $\hat\bs_\calS$, must be compared against the target to compute the loss and perform backpropagation. However, since we assume the target is not available, we use the pseudo target computed elsewhere, i.e., using the teacher model. 

This process falls in the category of the KD framework in which a student model is optimized using a teacher model's prediction \cite{HintonG2015arxiv}. In our PSE context, we employ a large pre-trained teacher model $\calT(\cdot)$ whose predicted clean utterance serves as the target to compute the student model's loss. Both student and teacher models are initialized with pre-trained generalist SE models. During test-time, the student model is optimized as: $\argmin_{\bTheta_\calS}  \calL(\hat\bs_\calT||\calS(\bx; \bTheta_\calS))$,
% \begin{equation}
%     \label{eqn:pse}
%     \argmax_{\bm{\Theta}_S}\; \text{SI-SDR}(\hat{\bm{s}}_T,\hat{\bm{s}}_S) - \text{SI-SDR}(\hat{\bm{s}}_T,\bm{x})
% \end{equation}
where $\hat{\bm{s}}_{\calT}$ is the estimates of clean speech signals obtained from the teacher model and $\bTheta_\calS$ are trainable parameters of the student model. 
 
The teacher's estimates $\hat{\bm{s}}_{\calT}$ are only approximations of ground-truth targets $\bs$, and can contain denoising artifacts \cite{XuY2014ieeespl}. However, under a zero-shot condition, we assume having these synthesized pseudo targets is better than nothing. Hence, the performance of the fine-tuning results depends on the quality of $\hat{\bm{s}}_{\calT}$. To this end, we employ relatively large models that surely outperform the student models on the test signals, i.e.,  $\calL(\bm{s}||\hat{\bm{s}}_\calT)<\calL(\bm{s}||\hat{\bm{s}}_\calS)$. Thus, we hypothesize that the student will still learn from these imperfect targets and improve its test-time SE performance. 
 
\begin{figure*}[t]
        \centering
        \begin{subfigure}[b]{0.47\textwidth}
            \centering \includegraphics[width=\textwidth]{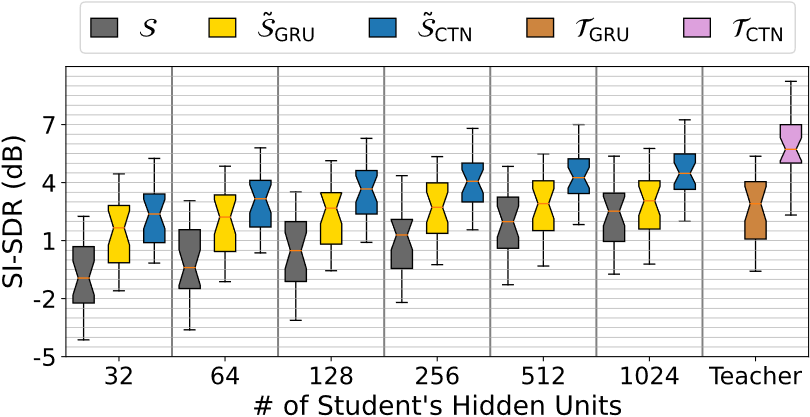}
            \caption[snr-5]%
            {{\small -5 dB Mixture SNR}}    
            \label{fig:snr-5}
        \end{subfigure}
        \hfill
        \begin{subfigure}[b]{0.47\textwidth}  
            \centering  \includegraphics[width=\textwidth]{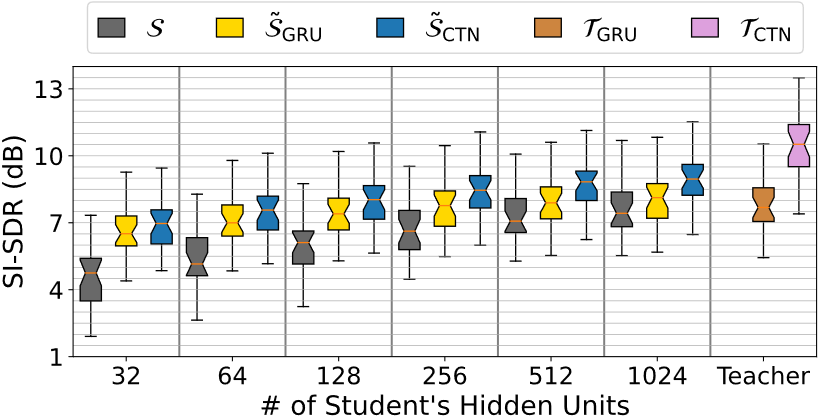}
            \caption[snr0]%
            {{\small 0 dB Mixture SNR}}    
            \label{fig:snr0}
        \end{subfigure}
        \vskip\baselineskip
        \begin{subfigure}[b]{0.47\textwidth}   
            \centering  \includegraphics[width=\textwidth]{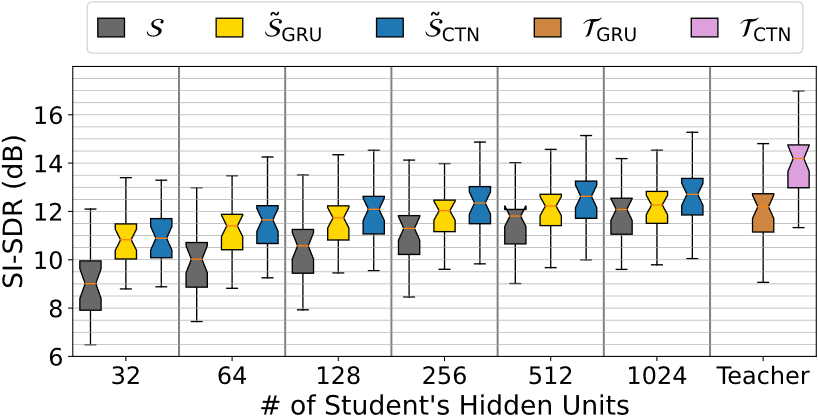}
            \caption[snr5]%
            {{\small 5 dB Mixture SNR}}    
            \label{fig:snr5}
        \end{subfigure}
        \hfill
        \begin{subfigure}[b]{0.47\textwidth}   
            \centering  \includegraphics[width=\textwidth]{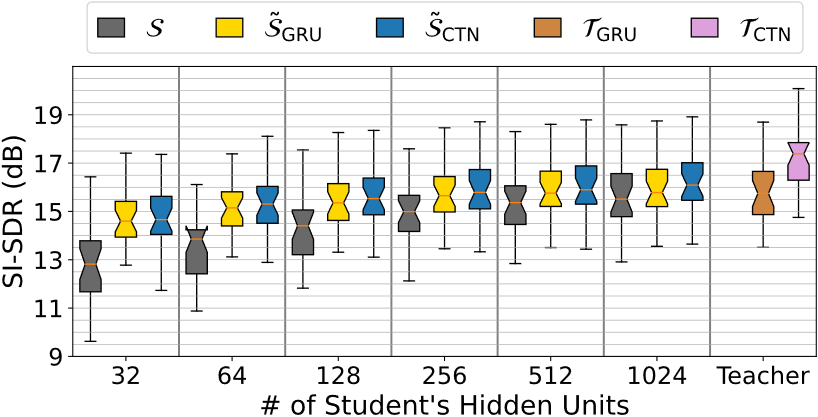}
            \caption[snr10]%
            {{\small 10 dB Mixture SNR}}    
            \label{fig:snr10}
        \end{subfigure}
        \caption[something]
        {\small Comparison of SE performances from pre-trained generalists against personalized specialists under various mixture SNR levels. Student models are initialized as 2-layered GRU generalists. Teacher models are provided as references.} 
        \label{fig:box}
    \end{figure*}

\section{Experiments}
\label{sec:experiments}

\subsection{The Datasets}
\label{sec:data}

% TODO: FT

For pre-training SE models, we used clean speech recordings from the LibriSpeech corpus \cite{PanayotovV2015Librispeech} and noise recordings from the MUSAN dataset \cite{SnyderD2015MUSAN}. We used Librispeech's \texttt{train-clean-100} and \texttt{dev-clean} subset for training and validation, which we denote as $\mathbb{G}_\text{tr}$ and $\mathbb{G}_\text{va}$ respectively. We split MUSAN's \texttt{free-sound} subset into training and validation partitions at 80:20 ratio, denoted as $\mathbb{N}_\text{tr}$ and $\mathbb{N}_\text{va}$ respectively. This exposes the generalist models to up to 251 speakers and 843 noise recordings during training. The noisy mixtures are obtained by adding the noise to speech signals at random input SNR levels uniformly chosen between -5 and 10 dB. 

For fine-tuning, we used 30 speakers from Librispeech's \texttt{test-clean} and noise from the WHAM! corpus \cite{WichernG2019wham} whose samples are recorded in 44 different locations. From these sets, we can create up to $K=30$ unique test environment by assigning a unique noise location to each speaker. 
% We synthesize unique test-time environments for each speaker by mixing utterances from each speaker with noise samples from a single location. 
Given a test environment index $k\in\{1,\ldots,K\}$, we extract clean speech signals from the $k$-th speaker $\mathbb{S}^{(k)}$ and add noises from $k$-th location $\mathbb{M}^{(k)}$. For each test environment, $\mathbb{S}^{(k)}$ and $\mathbb{M}^{(k)}$ are split into separate sets: the partitions are approximately 5, 1, and 1 minutes of clean speech, which we denote by $\mathbb{S}^{(k)}_\text{ft}$, $\mathbb{S}^{(k)}_\text{va}$ and $\mathbb{S}^{(k)}_\text{te}$. The noise datasets are prepared similarly: $\mathbb{M}^{(k)}_\text{ft}$, $\mathbb{M}^{(k)}_\text{va}$ and $\mathbb{M}^{(k)}_\text{te}$. We use $\mathbb{S}^{(k)}_\text{ft}$ and $\mathbb{M}^{(k)}_\text{ft}$ to fine-tune the student model via the KD process, where the teacher model's denoising results on the mixture of the pair are used as the pseudo target. $\mathbb{S}^{(k)}_\text{va}$ and $\mathbb{M}^{(k)}_\text{va}$ are mixed up to validate the student model during fine-tuning, mainly to prevent overfitting. Finally, we set aside $\mathbb{S}^{(k)}_\text{te}$ and $\mathbb{M}^{(k)}_\text{te}$ to test the final performance of the fine-tuned PSE system. 

When we simulate various test conditions, the noise and speech sources are mixed under four different input SNR levels (i.e. -5 dB, 0 dB, 5 dB and 10 dB) and used them for fine-tuning, validation, and testing. All audio files are loaded at 16 kHz sampling rate and standardized to have a unit-variance. 

% Short-time Fourier transform (STFT) converts the signal into the time-frequency (TF) domain with a Hann window of size 1024 samples and a 75\% overlap. 

% The duration of samples in $\mathbb{S}^{(k)}_\text{te}$ and $\mathbb{M}^{(k)}_\text{te}$ 10-seconds, extracted from a single utterance and noise recording respectively. For the fine-tuning and validation speech and noise sets, all audio samples are split into 1-second snippets. All samples within each dataset are standardized to have unit-variance.

\subsection{Models}
\label{sec:models}

Most of our SE models are based on the uni-directional gated recurrent unit (GRU) architecture.
% \cite{ChoK2014arxiv}. 
We use frequency-domain representations of the mixture signals obtained through the short-time Fourier transform (STFT) as inputs to the SE models. Spectrograms are generated using the STFT with a frame size of 1024 samples, a hop size of 256 samples, and a Hann window of 1024 samples. A dense layer transforms the GRU's output into the ideal ratio masks (IRM), which contains the probability of the TF bin belonging to the target source \cite{NarayananA2013icassp}. The denoising mask is applied element-wise to the mixture spectrogram, then transformed back to the time-domain signal $\hat{\bs}$ through inverse STFT. Finally, we use negative scale-invariant signal-to-noise ratio (SI-SNR) as the loss function \cite{LeRouxJL2018sisdr}. 

While the GRU architecture for the student models is fixed with two hidden layers, we vary their hidden units from 32 to 1024 to verify the impact of PSE on the different architectural choice of the student models. Meanwhile, as for the teacher model, we use a $3\times 1024$ GRU architecture, which is large enough  to outperform the students. In addition to the large GRU architecture, we also employ ConvTasNet (CTN) \cite{LuoY2019conv-tasnet} as an alternative teacher model.  Since the CTN teacher outperforms the GRU model due to its structural advantage, we can confirm the impact of the teacher's performance on the KD-based PSE. The CTN model is configured using implementation available in Asteroid's source separation toolkit \cite{ParienteM2020asteroid}. Same architecture as reported in \cite{LuoY2019conv-tasnet} is adopted (i.e. 8 convolutional blocks and 3 repeats with global layer normalization) and trained on a single-speaker speech enhancement task. The model architectures, their respective number of parameters, and the multiplier-accumulator (MAC) operation counts are shown in Table \ref{tab:cpl}. Note that CTN is not the largest model but it requires extensive MAC operations. 

Both teacher models are pre-trained as generalist SE models using noisy mixtures generated from adding $\mathbb{G}_\text{tr}$ and $\mathbb{N}_\text{tr}$. We select the best models using early-stopping determined from validation computed using $\mathbb{G}_\text{va}$ and $\mathbb{N}_\text{va}$. During pre-training, the clean speech dataset are used as the ground-truth targets. 

% The student models for test-time PSE are use 2-layered GRU-based student models with hidden unit sizes of [32, 64, 128, 256, 512, 1024]. We test each of the students on a teacher model with larger architecture, 3-layered GRU with 1024 hidden units and a ConvTasNet model. 

The student models are fine-tuned using mixtures of $\mathbb{S}_\text{ft}$ and $\mathbb{M}_\text{ft}$. Their best models are determined through validation on the set-aside validation set $\mathbb{S}_\text{va}$ and $\mathbb{M}_\text{va}$. Finally, we test the fine-tuned models on the mixture of $\mathbb{S}_\text{te}$ and $\mathbb{M}_\text{te}$, which have not been exposed to any of the pre-training and fine-tuning processes. 

% HERE: The models are tested on STE MTE for score reporting...

% \begin{equation}
% \begin{split}
% & \hat{\bm{s}}_\calT\\
% & \hat{\bm{s}}_\calS\\
% \end{split}
% \end{equation}

% During the fine-tuning phase, the speech dataset are used as the ground-truth targets (Eqn. \ref{eqn:pse}).

% All networks during both pre-training and fine-tuning phases are optimized using the Adam optimizer \cite{KingmaD2015adam}. 

\renewcommand{\arraystretch}{1.}
% \hspace*{-2cm}
\begin{table}[t!]
\centering
\caption{Complexity of student and teacher models in MACs and number of parameters. MACs are computed given 1-second inputs.}
% \vspace{1mm}
% \hspace{-5mm}
\begin{tabular}{c|c|c|c}
\hline
\multicolumn{2}{c|}{Models} & MACs (G) & Param. (M)\\
\hline\hline
\multirow{6}{*}{Student} & GRU (2$\times$32) & 0.010 & 0.08 \\
\cline{2-4}
& GRU (2$\times$64) & 0.011 & 0.17 \\
\cline{2-4}
& GRU (2$\times$128) & 0.026 & 0.41 \\
\cline{2-4}
& GRU (2$\times$256) & 0.071 & 1.12 \\
\cline{2-4}
& GRU (2$\times$512) & 0.216 & 3.42 \\
\cline{2-4}
& GRU (2$\times$1024) & 0.729 & 11.55 \\
 \hline
 \multirow{2}{*}{Teacher} & GRU (3$\times$1024) & 1.126 & 17.85 \\
\cline{2-4}
& ConvTasNet \cite{LuoY2019conv-tasnet} & 9.831 & 4.92 \\
 \hline
\end{tabular}
\label{tab:cpl}
\end{table}

\section{Experimental Results and Discussions}
\label{sec:disc}

% Improvements for all cases on average. Sometimes better than teacher
The box plots in Figure \ref{fig:box} show the SE performances for various models under environments synthesized from different noise level conditions. The results are shown for pre-trained and fine-tuned student models and the teacher model as the reference. Here, we introduce new notations to distinguish the two teacher models with the GRU and CTN architectures, $\calT_\text{GRU}$ and $\calT_\text{CTN}$, respectively. In addition, we also denote the fine-tuned students models differently from the pre-trained initial model $\calS$ and add the subscript to indicate what it learns from: $\tilde{\calS}_\text{GRU}$ and $\tilde{\calS}_\text{CTN}$, respectively. Hence, each box that represents one of the generalist models, $\calS$, $\calT_\text{GRU}$, and $\calT_\text{CTN}$, is an average SI-SDR performance of the system on all 30 unique test environments. On the other hand, a box for one of the specialist architectures, $\tilde{\calS}_\text{GRU}$ and $\tilde{\calS}_\text{CTN}$, is an average performance of 30 different personalized models on the 30 test conditions, applied respectively.  

Our proposed PSE framework improves all pre-trained student models under all noise conditions, i.e., $\tilde{\calS}_\text{GRU}$ and $\tilde{\calS}_\text{CTN}$ results are always better than the $\calS$ results on average. In addition, we also observe that the personalized models learned from the CTN teacher, $\tilde{\calS}_\text{CTN}$, always outperform their corresponding ones fine-tuned using the GRU teacher, $\tilde{\calS}_\text{GRU}$. Given that each student pair in comparison are stemmed from the same pre-trained GRU model, it showcases that the quality of the teacher model's performance is related to the performance of fine-tuning. It is also noticeable that the structural discrepancy between the student and teacher, i.e., $\tilde{\calS}_\text{CTN}$ (a GRU) and ${\calT}_\text{CTN}$ (a CTN), is not an issue. 

The smaller student models show more significant improvements via PSE. Hence, it verifies that PSE is a model compression method, because a smaller personalized model can compete with a large generalist (e.g. 2 $\times$ 32 $\tilde{\calS}_\text{CTN}$ vs. 2 $\times$ 1024 ${\calS}_\text{GRU}$ for -5 dB mixture SNR as in Figure \ref{fig:snr-5}). According to Table \ref{tab:cpl}, a personalized 2 $\times$ 32 specialist saves 11.47M parameters and 719M MACs compared to a 2 $\times$ 1024 generalist (for 1-second inputs). Likewise, instead of increasing generalists' architectures for better generalization capabilities, it is more advantageous to personalize the models. 

% Different teacher conditions (size, type).
When the teacher model is better than the student by only a small margin, personalized student models are even able to outperform the relative teacher model, i.e., $\tilde{\calS}_\text{GRU}$ (2 $\times$ 1024) vs. ${\calT}_\text{GRU}$ (3 $\times$ 1024). We believe it is because of the student model's dedicated exposure to the test-time environment during finetuning. 

%% Logistics / concluding remarks
Personalization could potentially worsen the student's generalization performance on other unseen speakers and noise types, which can be problematic if the surrounding changes. However, the student can always update again to reflect any changes in the deployed environment. We envision a scenario where the fine-tuning procedure can be done on the cloud, where the residing teacher model updates the small student model. To this end, the small student model needs to be transferred from the cloud server to the user device, which may not be burdensome given its small size. The cloud computing option is also convenient, as the finetuning step do not need to wait for the teacher model to denoise the test signals, which is an energy- and time-consuming process to be conducted in the small device. Likewise, frequent updates to the student does not become burdensome for the device. Since our framework is simple, we expect our framework to provide improvements under different data or loss functions, and even be applicable to other domains. 

% Maybe include the four plots with gray and red lines since have space. 

\section{Conclusion}
\label{sec:conc}

% TODO: figure again with more ticks and seeds.

% After fine-tuning the smaller student models, we can outperform larger models while occupying lesser memory space and reducing compupational complexity; thereby saving energy and carbon footprint of the deployed models. 

In this paper, we proposed a simple zero-shot learning framework that utilizes knowledge distillation to fine-tune a speech enhancement model during test-time, which we call personalization. By utilizing the teacher's estimates as the targets, which otherwise do not exist during the test time, we showed that the student model's performance greatly improves on a specific test-time speaker and the acoustic environment. Since our small personalized student model can give superior performances to large generalist models, we claim that the knowledge distillation-based fine-tuning method provides another mode of model compression that does not sacrifice performance. Our framework is flexible as it can employ heterogeneous model architectures within a teacher-student pair. Our zero-shot personalization procedure does not require any ground-truth clean speech signals from the test-time user, making it more mindful about users' privacy. Finally, we envision that PSE can be a solution to improving the model's performance on the user groups that are underrepresented in the training set. The source codes and sound examples are available at: \hyperlink{https://saige.sice.indiana.edu/research-projects/KD-PSE}{https://saige.sice.indiana.edu/research-projects/KD-PSE}

% \section{ACKNOWLEDGMENT}
% \label{sec:ack}

\newpage 
% -------------------------------------------------------------------------
% Either list references using the bibliography style file IEEEtran.bst
\bibliographystyle{IEEEtran}
\bibliography{main}
% \bibliography{mjkim,refs21}

\end{sloppy}
\end{document}